\documentclass[manuscript]{aastex}
\usepackage{aas_macros,url,natbib,amssymb,amsmath,CJK}

\shorttitle{209P/LINEAR and the Camelopardalid Meteor Shower}
\shortauthors{Ye et al.}

\begin{document}
\begin{CJK*}{UTF8}{gbsn}

\title{When Comets Get Old: A Synthesis of Comet and Meteor Observations of the Low Activity Comet 209P/LINEAR}

\author{Quan-Zhi Ye (叶泉志)}
\affil{Department of Physics and Astronomy, The University of Western Ontario, London, Ontario N6A 3K7, Canada}
\email{qye22@uwo.ca}

\author{Man-To Hui (许文韬)}
\affil{Department of Earth, Planetary and Space Sciences, University of California at Los Angeles, 595 Charles Young Drive East, Los Angeles, CA 90095-1567, U.S.A.}

\author{Peter G. Brown, Margaret D. Campbell-Brown, Petr Pokorn\'{y}, and Paul A. Wiegert}
\affil{Department of Physics and Astronomy, The University of Western Ontario, London, Ontario N6A 3K7, Canada}
\affil{Centre for Planetary Science and Exploration, The University of Western Ontario, London, Ontario N6A 5B8, Canada}

\and

\author{Xing Gao (高兴)}
\affil{No. 1 Senior High School of \"{U}rumqi, \"{U}rumqi, Xinjiang, China}

\begin{abstract}
It is speculated that some weakly active comets may be transitional objects between active and dormant comets. These objects are at a unique stage of the evolution of cometary nuclei, as they are still identifiable as active comets, in contrast to inactive comets that are observationally indistinguishable from low albedo asteroids. In this paper, we present a synthesis of comet and meteor observations of Jupiter-family comet 209P/LINEAR, one of the most weakly active comets recorded to-date. Images taken by the Xingming 0.35-m telescope and the Gemini Flamingo-2 camera are modeled by a Monte Carlo dust model, which yields a low dust ejection speed ($1/10$ of that of moderately active comets), dominance of large dust grains, and a low dust production of $0.4~\mathrm{kg \cdot s^{-1}}$ at 19~d after the 2014 perihelion passage. We also find a reddish nucleus of 209P/LINEAR that is similar to D-type asteroids and most Trojan asteroids. Meteor observations with the Canadian Meteor Orbit Radar (CMOR), coupled with meteoroid stream modeling, suggest a low dust production of the parent over the past few hundred orbits, although there are hints of a some temporary increase in activity in the 18th century. Dynamical simulations indicate 209P/LINEAR may have resided in a stable near-Earth orbit for $\sim 10^4$~yr, which is significantly longer than typical JFCs. All these lines of evidence imply that 209P/LINEAR as an aging comet quietly exhausting its remaining near surface volatiles. We also compare 209P/LINEAR to other low activity comets, where evidence for a diversity of the origin of low activity is seen.
\end{abstract}


\section{Introduction}

Dormant comets are comets that have depleted their volatiles in the near surface layers but may still possess an ice-rich interior. It is not easy to study these objects directly, as their optical properties are indistinguishable from those of some of their asteroidal counterparts. Dormant comets among the population of near-Earth objects (NEOs) are particularly interesting, as they may have a significant contribution to Earth's history. It has been suggested $\sim 10\%$ of NEOs had their origins as Jupiter-family Comets or JFCs \citep[e.g.][]{2002Icar..159..358F,2008Icar..194..436D}.

The dynamical lifetime of common JFCs is about $10^5$~yr \citep{Levison1994}. The physical lifetime of kilometer-sized JFCs, however, is estimated to be only a few $10^3$~yr \citep[e.g.][]{2009Icar..203..140D}. It is therefore evident that a typical JFC, presuming it does not fragment or split, would spend most of its time as a dormant comet. The details of the active-dormancy transition remain nebulous, but classical understanding of cometary evolution argues that the transition might include a period of low or intermittent cometary activity, possibly due to the buildup of dust mantles on the surface \citep[c.f.][]{Jewitt2004d}. Hence, it is natural to speculate that some weakly active comets may be active-dormancy transitional objects. From an observer's perspective, these objects are at a unique stage of the evolution of cometary nuclei, as they are still observationally identifiable as physical comets, as opposed to completely dormant comets that are indistinguishable from low albedo asteroids.

We define a low activity comet as a comet where the absolute total magnitude, $M_1$, is higher (fainter) than the absolute magnitude of a dark asteroid (defined by V-band geometric albedo $p_\mathrm{v}=0.1$) of equivalent effective body (nucleus) size. The physical implication of this definition is that the cometary activity is so low, that the comet would be recognized as a dark asteroid ($p_\mathrm{v}<0.1$) if extended cometary features are unresolvable to an observer. Mathematically, the definition can be expressed as

\begin{equation}
M_1 > 16.6 - 5\log{\left( \frac{R_\mathrm{N}}{1~\mathrm{km}} \right)}
\end{equation}

\noindent where $R_\mathrm{N}$ is the effective nucleus radius. Among the 121 comets with constrained nucleus sizes\footnote{The nucleus sizes of these 121 comets are extracted from the JPL Small-Body Database (\url{http://ssd.jpl.nasa.gov/sbdb_query.cgi}) on 2015 June 3.}, we find 9 comets meeting our definition of low activity comets (Table~\ref{tbl:wc}) of which 8 are near-Earth JFCs.

What are the nature and the origins of these comets? To answer this question, we need to look at their physical and dynamical properties. In particular, we note four of these comets -- namely 209P/LINEAR, 252P/LINEAR, 289P/2003 WY25 (Blanpain) and 300P/2005 JQ5 (Catalina) -- can produce meteor showers currently observable at Earth. Meteor showers are caused by cometary dusts ejected in past orbits of the parent, therefore meteor observations have the potential of enhancing our understanding of the physical history of the parent, as demonstrated in the investigation of the present and past activity of 55P/Tempel-Tuttle \citep[e.g.][]{Yeomans1981,Brown1999a} and a couple of potential dormant comets \citep[e.g.][]{Babadzhanov2012,Kokhirova2015}.

In this paper, we focus on one particular comet in our list, 209P/LINEAR. 209P/LINEAR is among the most weakly active comets ever recorded \citep[e.g.][]{2014CBET.3881....2S,Ishiguro2015} and is associated with a new meteor shower, the Camelopardalids \citep[e.g.][]{Jenniskens2014,Madiedo2014l}. What makes 209P/LINEAR ideal in studying cometary dormancy transition is (1) the close approach to the Earth of the comet during its 2014 perihelion passage, reaching $\sim 0.05$~AU from the Earth where it had brightened to $V\sim 11$~magnitude; and (2) the simultaneous encounter of a series of dust trails produced by the comet in its past orbits. These two events provide a rare opportunity to look at a potential comet-asteroid transitional object from two complementary approaches. Therefore, we observe 209P/LINEAR itself (\S2) as well as the associated meteor activity (\S3) to characterize the current state and recent history of the comet's activity. The observations are coupled with the results from numerical simulations to understand the nature and origin of 209P/LINEAR (\S4). We also discuss the implication of our results to the state of to other low activity comets through the examination of 209P/LINEAR.

\section{The Comet}

\subsection{Observation}

Imaging observations were conducted with three facilities at three different epochs. The observations and reduction procedures are summarized below and tabulated in Table~\ref{tbl:obs}.

\begin{enumerate}
 \item Gemini North + Gemini Multi-Object Spectrographs (GMOS) camera at 2014 April 9.25 UT. This is a single frame taken as a snapshot observation. The observation was conducted relatively early in the active phase of 209P, making it suitable for examining the initial activation of the comet.
 \item The 0.35-m telescope + QHY-9 camera at Xingming Observatory on 2014 May 18.75 UT. Around this date, the viewing geometry was favorable for separating dust of different sizes and emission epochs. The observation was conducted without filters and was processed using standard procedures (bias and dark frame subtraction, flat frame division).
 \item Gemini South + Flamingo-2 (F-2) camera on 2014 May 25.94 UT. Around this date, the Earth was close to the comet and was near the orbital plane of the comet. The observation was conducted in the $K_\mathrm{s}$ band, with 15~s of exposure of each frame. The telescope was nodded in the direction perpendicular to the tail axis, to avoid contamination from the tail signal at the sky subtraction stage. As the comet was moving at a fast rate of $\sim 18''$/min (or 25~pix per frame), we opted for the non-guided non-sidereal tracking mode to avoid frequent changes of guide stars. Because of this, a small fraction ($<5\%$) of frames suffer from poor tracking and are discarded. At the end, a total of 41 frames were useful for later analysis. The data reduction is performed with the Image Reduction and Analysis Facility (IRAF) supplied by Gemini.
\end{enumerate}

\subsection{Results and Analysis}

\subsubsection{Start of Cometary Activity and General Morphology}
\label{sec:start_of_activity}

\begin{figure}
\includegraphics[width=0.5\textwidth]{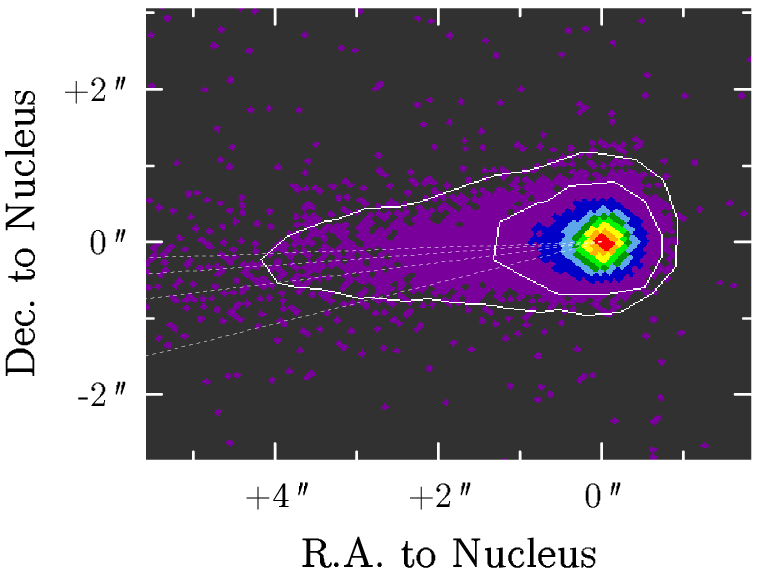}
\caption{The 2014 Apr. 9 GMOS-N image (stretched in logarithm scale) superimposed with the synchrone model. The ages of the synchrone lines (dashed lines) are (in counterclockwise order) 10, 25, 50 and 100~d respectively. The oldest visible dust was released at $\tau \sim 50$~d, appropriate to late Feb. 2014.}
\label{fig:syn}
\end{figure}

Previous researches \citep{Hergenrother2014,Ishiguro2015} found that the activity of 209P/LINEAR started at a small activation distance of $r_\mathrm{H}=1.4$~AU. With the GMOS image, we conduct an independent check of the start time of activity of 209P/LINEAR. This is done by comparing the surface brightness profile to a synchrone model \citep{Finson1968}. We estimate the start of activity occurred no later than late February 2014 or a lead time of $\tau \sim 50$~d, where 209P/LINEAR was at $r_\mathrm{H}=1.4$~AU (Figure~\ref{fig:syn}). This is in agreement with previous results.

\begin{figure*}
\includegraphics[width=\textwidth]{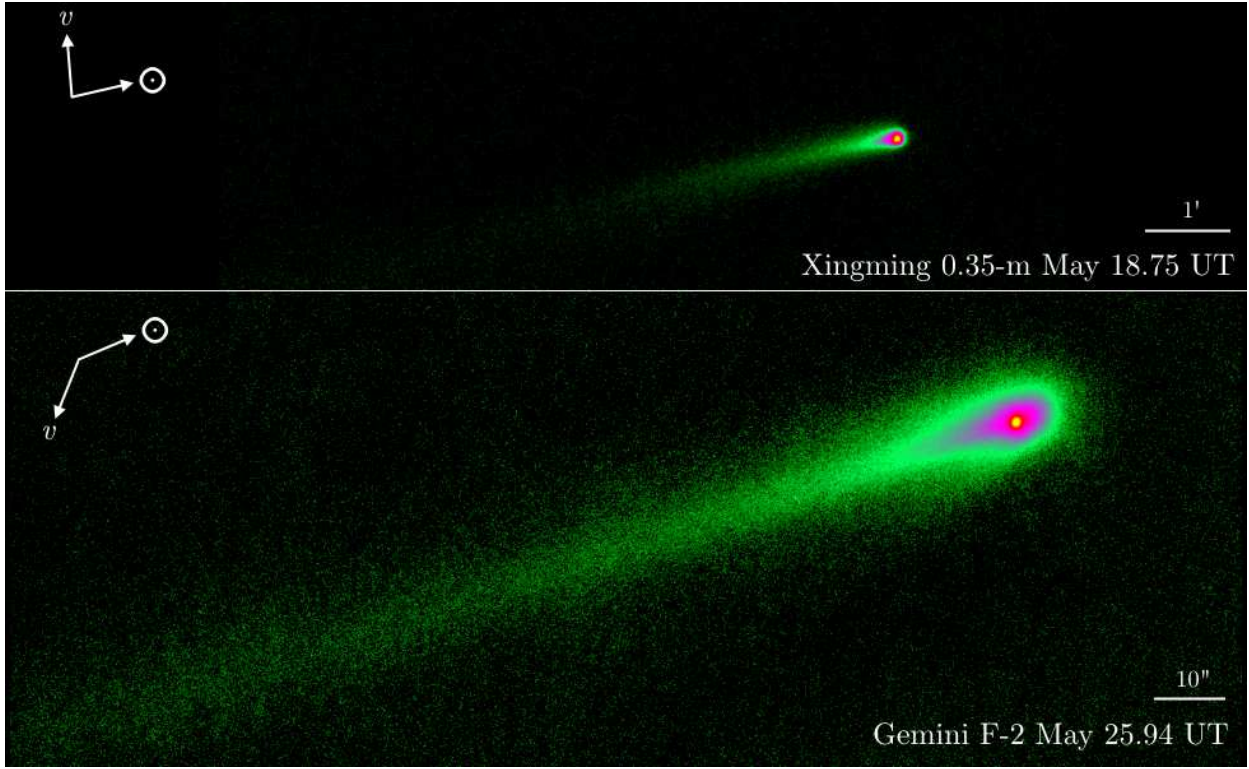}
\caption{Composite images of 209P/LINEAR taken by Xingming 0.35-m telescope and Gemini Flamingo-2 on 2014 May 18 and 25. The images are stretched in asinh scale and are rotated to have north-up east-left.}
\label{fig:img}
\end{figure*}

Composite images taken by Xingming 0.35-m telescope and Gemini F-2 on May 18 and 25 are shown as Figure~\ref{fig:img}. In the optical image from Xingming, 209P/LINEAR showed a symmetric coma measured $6-7''$ (or about $50\%$ larger than mean Full-Width-Half-Maximum or FWHM of background stars) in size and a mostly straight dust tail extended beyond the field of view. In the near infrared image from F-2, the nucleus, with the same FWHM compared to background stars, is clearly separated from the coma. The coma is significantly elongated along the Sun-comet axis, with the sunward side extending $\sim5''$ or $\sim 230$~km towards the solar direction.

\subsubsection{Modeling the Dust}
\label{sec:dust_model}

To understand the dust properties, we model the observations using a Monte Carlo dust model evolved from the one used in \citet{2014ApJ...787..115Y}. The dynamics of the cometary dust are determined by two parameters: the ratio between radiation pressure and solar gravity, $\beta_\mathrm{rp}=5.7\times10^{-4}/(\rho_\mathrm{d} a_\mathrm{d})$, where $\rho_\mathrm{d}$ the bulk density of the dust and $a_\mathrm{d}$ the diameter of the dust, both in SI units \citep{Wyatt1950e,Burns1979ai}; and the initial ejection velocity of the dust. The latter is found following the philosophy of the physical model proposed by \citet{Crifo1997a}, is defined as

\begin{equation}
v_\mathrm{ej} = V_0 \beta_\mathrm{rp}^{1/2} \cos z \cdot \nu
\end{equation}

\noindent where $V_0$ is the mean ejection speed of a dust particle of $\beta_\mathrm{rp}=1$, $z$ is the local solar zenith angle, and $\nu$ follows a Gaussian probability density function:

\begin{equation}
P(\nu) = \mathcal{N}(1,\sigma_\nu^2)
\end{equation}

\noindent where $\sigma_\nu$ is the standard deviation of $\nu$. The $P(\nu)$ function heuristically accounts for the variable shape and cross-section of the cometary dust that affects the radiation force impulse experienced by the dust.

We assume the dust size follows a simple power-law with a differential size index of $q$. Therefore, the dust production rate is expressed as

\begin{equation}
N(r_\mathrm{H}, a_\mathrm{d}) \mathrm{d} a_\mathrm{d} = N_0 \left( \frac{r_\mathrm{H}}{1~\mathrm{AU}} \right)^{-k} \left( \frac{a_\mathrm{d}}{1~\micron} \right)^{-q} \mathrm{d} a_\mathrm{d}
\end{equation}

\noindent where $N_0$ is the mean dust production rate of $1~\micron$ particles and $r_\mathrm{H}$ is the heliocentric distance at which the dust is released. We use $k=4$ following the canonical comet brightening rate \citep[c.f.][]{Everhart1967}.

We assume the observed flux is solely contributed by scattered light from the dust particles released in the current perihelion passage, and set the start epoch of dust emission to 2014 Feb. 18 ($\tau=50$~d) as found in \S~\ref{sec:start_of_activity}. Simulated particles are symmetrically released around the comet-Sun axis line at the sunlit side. The size distribution is set to the interval of the free parameter $\beta_\mathrm{rp,max}$ (i.e. the lower limit of dust size) to an upper size limit constrained by the escape speed $v_\mathrm{esc}=\sqrt{2GM_\mathrm{N}/R_\mathrm{G}}$, where $M_\mathrm{N}=\frac{4}{3}\pi R_\mathrm{N}^3 \rho_\mathrm{N}$ is the total mass of the nucleus, $\rho_\mathrm{N}=500~\mathrm{kg \cdot m^{-3}}$ the bulk density of the nucleus, $R_\mathrm{N}=1.35$~km the effective nucleus radius \citep{Howell2014v}, and $R_\mathrm{G}=10R_\mathrm{N}$ the characteristic distance that gas drag become negligible \citep{Gombosi1986f}. A modified MERCURY6 package \citep{Chambers1999t} is used to integrate particles from the start epoch to the observation epoch using the 15th order RADAU integrator \citep{Everhart1985}. Gravitational perturbations from the eight major planets (the Earth-Moon system is represented by a single mass at the barycenter of the two bodies), radiation pressure and Poynting-Robertson effect are included in the integration. The orbital elements of 209P/LINEAR are extracted from the JPL small body database elements 130 (\url{http://ssd.jpl.nasa.gov/sbdb.cgi}) and are listed in Table~\ref{tbl:mdl-orb} together with other parameters used for the model. The resulting image is convolved with a 2-dimensional Gaussian function (with FWHM equals to the FWHM of the actual images) to mimic observational effects such as the instrumental point spread effect and atmospheric seeing.

\begin{figure}
\includegraphics[width=0.5\textwidth]{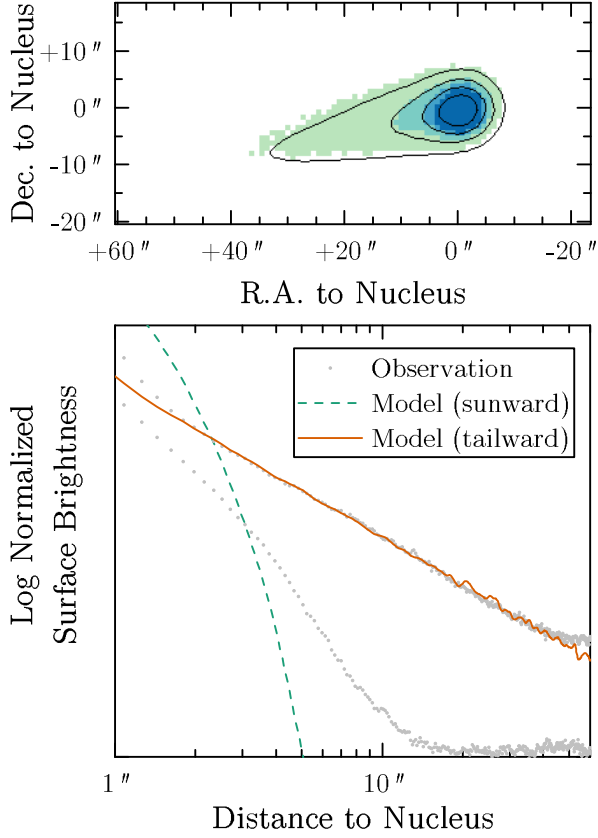}
\caption{Observed (colored pixels) and modeled (contours) surface brightness profiles for the Xingming image (upper figure) and the Gemini F-2 image (lower figure; the sunward data is shifted downwards for clarity). The surface brightness profiles are normalized to the pixel intensity 3~FWHMs behind the nucleus along the Sun-comet axis to avoid contamination from the nucleus signal. The mean best model for both the Xingming and the Gemini F-2 images has $\beta_\mathrm{rp,max}=0.004$ to 0.005, $V_0=40~\mathrm{m \cdot s^{-1}}, q=3.8$ and $\sigma_\nu=0.3$.}
\label{fig:img-mdl}
\end{figure}

We first model the May 18 Xingming image with the following procedure. First, we select the pixels $>3\sigma$ from the background with $\sigma$ the standard deviation of all the pixels in the image. Observed and modeled surface brightness profiles are then normalized to 3~FWHMs beyond the nucleus along the Sun-comet axis, with the region within 1~FWHM from the nucleus being masked out, as the signal from the nucleus may contaminate the central condensation. The degree of similarity of the two profiles is then evaluated using the normalized error variance (NEV), under a polar coordinate system centered at the nucleus with angular resolution of $1^\circ$:

\begin{equation}
\label{eq:nev}
\mathrm{NEV} = \frac{1}{n} \sum_{i=1}^n \frac{\sqrt{(\mathbf{M}_i - \mathbf{O}_i)^2}}{\mathbf{O}_i}
\end{equation}

\noindent where $n$ is the number of pixels above $3\sigma$ from the background, $\mathbf{M}_i$ and $\mathbf{O}_i$ are the pixel brightness from the modeled and observed brightness profile respectively. We set the tolerance level of NEV to 5\% in order to derive uncertainties of the model parameters. We then test a range of parameters as tabulated in Table~\ref{tbl:mdl}, which yields $\beta_\mathrm{rp,max}=0.005$, $V_0=40\pm10~\mathrm{m \cdot s^{-1}}, q=3.8\pm0.4$ and $\sigma_\nu=0.3\pm0.1$, shown as Figure~\ref{fig:img-mdl}). We find the dominance of larger dust in general agreement with previous results \citep[e.g.][]{2014MNRAS.437.3283Y,Younger2015}, except a steeper size distribution ($q=3.8$ vs. $q=3.25$) and a slightly lower ejection velocity ($v_\mathrm{ej}=1.5~\mathrm{m \cdot s^{-1}}$ vs. $v_\mathrm{ej}=2.5$ to $4.4~\mathrm{m \cdot s^{-1}}$ for millimeter-sized dust) comparing to the results from \citet{Ishiguro2015}. The ejection speed is about an order of magnitude lower than the one given by some classic ejection models \citep[e.g.][]{Jones1995q,Crifo1997a,Williams2001d} and is not much higher than the escape velocity ($v_\mathrm{esc}=0.2~\mathrm{m \cdot s^{-1}}$).

We then model the May 25 Gemini image. As the image was taken almost edge-on to the comet's orbital plane, dust particles at different sizes collapse onto the viewing plane, making it possible to collapse the image into a 1-dimensional profile without losing too much information. This comes with the benefit of simplifying subsequent analysis. Here we recognize that the orbital plane angle at the time of the observation ($+9.3^\circ$) was not really as small as those used in other studies (generally $<5^\circ$), therefore collapsing the image may result a loss in the resolution of the data, which should be reflected as an elevation in the uncertainties of the modeled parameters. However, we later see that the uncertainties of the best models of the May 25 Gemini image are comparable to that of the May 18 Xingming image (which was not collapsed). Hence, we conclude that the collapse of the image does not have a significant impact to our result.

\begin{figure*}
\includegraphics[width=\textwidth]{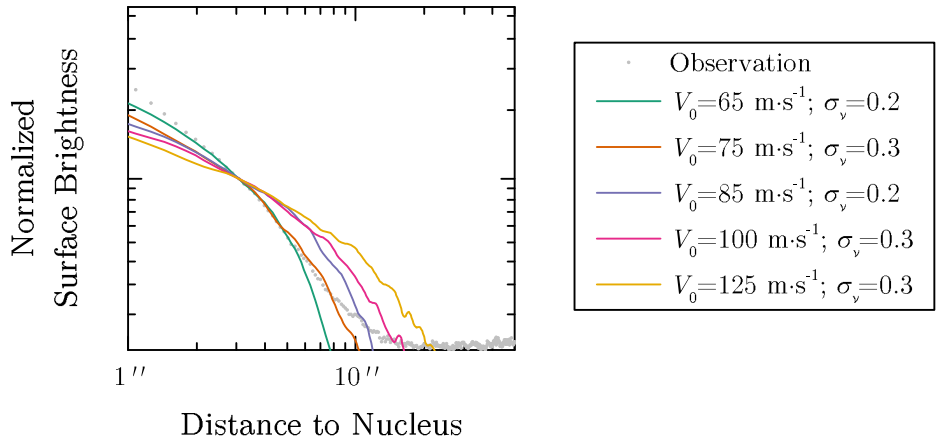}
\caption{Representative attempts to fit the sunward section of the coma in the Gemini F-2 image. The observed and modeled profiles are all normalized to 3~FWHMs away from the nucleus along the comet-Sun axis. These models have $q=-3.8$ and $\beta_\mathrm{rp,max}=0.004$.}
\label{fig:img-mdl-sunward}
\end{figure*}

We test the same range of parameters as listed in Table~\ref{tbl:mdl} to model the May 25 Gemini image. The observed and modeled surface brightness profiles are then integrated along the direction perpendicular to the orbital plane, and normalized to 3~FWHMs behind the nucleus along the Sun-comet axis. The goodness of the model is determined using Eq.~\ref{eq:nev}. We find $\beta_\mathrm{rp,max}=0.004$, $V_0=40\pm10~\mathrm{m \cdot s^{-1}}, q=3.8\pm0.4$ and $\sigma_\nu=0.3\pm0.1$, which is in good agreement with the parameters found from the May 18 Xingming image. However, we note that despite the fact that the fit at the tailward direction is good, the discrepancy between the modeled and the observed profile at the sunward direction is striking (Figure~\ref{fig:img-mdl}). Additional testing at the sunward-only section with the same test grid as Table~\ref{tbl:mdl} reveals no compatible dust model (Figure~\ref{fig:img-mdl-sunward}), which suggest a violation of the steady flow assumption we used for the model.

\subsubsection{Near Nucleus Environment}
\label{sec:near_nucleus}

\begin{figure*}
\includegraphics[width=\textwidth]{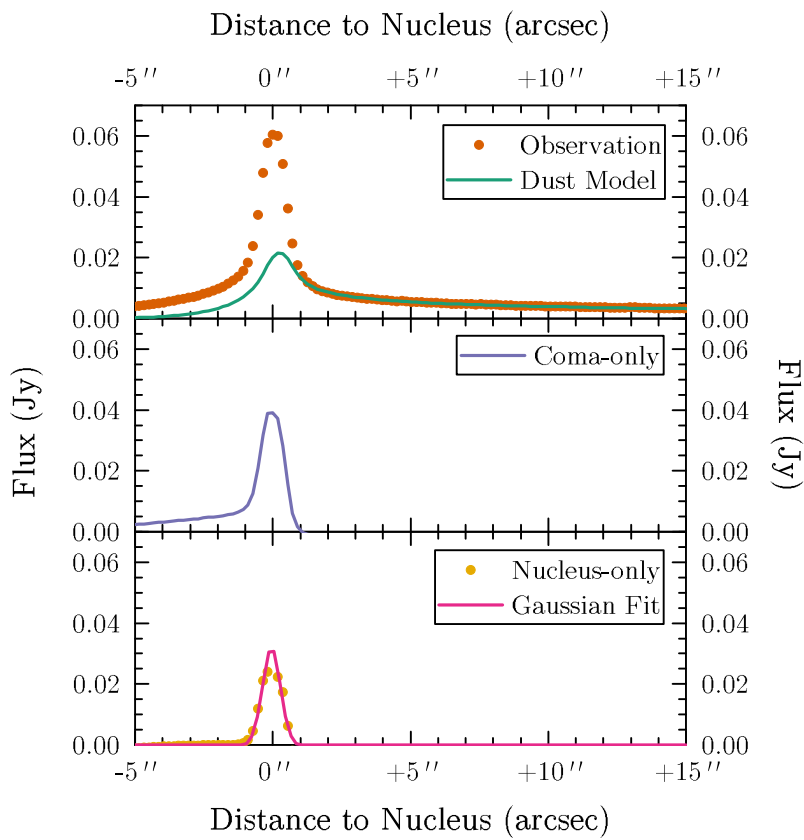}
\caption{Separation of the coma and nucleus signal based on the Gemini F-2 image. Upper figure: observed profile and modeled profile from the dust model. Middle figure: derived coma+nucleus profile by subtracting the observed profile with the modeled profile. Lower figure: nucleus-only profile, derived from subtracting the linear portion of the coma profile. The X-axis corresponds to the Sun-comet axis.}
\label{fig:mdl-mass}
\end{figure*}

To understand the physical properties of the non-steady coma, we separate the steady (i.e. the dust tail) and the non-steady component (i.e. the coma) in the surface brightness profile and calculate the flux for each of them. We first perform an internal absolute photometric calibration using the 2MASS stars in the image \citep{Skrutskie2006}. The selected calibration stars are at least $0.5'$ away from the tail axis to avoid contamination from the comet. We then correlate the observed profile to the modeled profile on the absolute scale. We subtract the modeled tail component (from \S~\ref{sec:dust_model}), and interpolate the linear portion of the coma to further isolate the nucleus signal (Figure~\ref{fig:mdl-mass}). This leaves the profiles of the steady tail and the non-steady coma calibrated to an absolute scale. By integrating these profiles, we derive the flux for the tail and the coma to be $\mathcal{F_\mathrm{tail}}=0.40$~Jy and $\mathcal{F_\mathrm{coma}}=0.04$~Jy respectively. The effective cross-section area for each component can be calculated by 

\begin{equation}
C_\mathrm{e} = \left( \frac{r_\mathrm{H}}{1~\mathrm{AU}} \right)^2 \frac{\pi \varDelta^2}{A_\lambda(\alpha)} \frac{F_\lambda}{F_{\odot, \lambda}}
\end{equation}

\noindent where $\varDelta$ is the geocentric distance, $F_\lambda$ and $F_{\odot, \lambda}$ is the flux of the component of interest and the Sun at the desired wavelength $\lambda$, which $F_{\odot, \lambda}=1.4\times10^{14}$~Jy for $K_\mathrm{s}$ band, and $A_\lambda(\alpha)$ the phase angle corrected geometric albedo.

For the tail component, the dust model gives a mean dust size $\bar{a}_\mathrm{d}=2\times10^{-4}$~m. By using $A_\lambda(0^\circ)=0.05$ and calculating the phase angle correction following the compound Henyey-Greenstein function \citep{Marcus2007a,Marcus2007b}, we derive $C_\mathrm{e}=5~\mathrm{km^2}$ and the corresponding dust mass $M_\mathrm{d}=\frac{4}{3} \rho_\mathrm{d} \bar{a}_\mathrm{d} C_\mathrm{e}=1\times10^6~\mathrm{kg}$. Considering the $r_\mathrm{H}$ dependency, the dust production rate at the observation epoch is calculated to be $\dot{M}_\mathrm{d}=0.4~\mathrm{kg \cdot s^{-1}}$, yielding a dust-water mass ratio of $\sim 1:2$ using the water production rate derived from narrow band observations \citep{2014CBET.3881....2S}. This is lower than other measurements \citep[e.g.][]{Kueppers2005,Rotundi2015} but is perhaps not unexpected, given the large scatter (within a factor of 10--100) of the dust-gas ratio among comets \citep{1995Icar..118..223A}. We also note the derived dust production is about an order of magnitude lower than the value derived by \citet{Ishiguro2015}, likely due to different model parameters (such as $\bar{a}$) used for the calculation.


On the other hand, the non-steady region extends no more than $\sim 2''$ behind the nucleus, corresponding to a mean lifetime of $\sim 1$~d appropriate to 10--100~$\micron$ particles. Interestingly, this is comparable to the mean lifetime of icy grains (purity $X_1\gtrsim0.9999$) of comparable sizes at $r_\mathrm{H}=1$~AU \citep[e.g.]{Hanner1981a,Beer2006c}, seemingly endorsing that the presence of an icy grain halo as a self-consistent explanation to the observation. However, we note that this hypothesis is not without problems: for dirtier icy grains such as $X_1=0.9$ grains, centimeter-sized grains would be required to survive to 1~d; we also note that icy grain halos are known to exist only on long period comets and hyperactive JFCs \citep[c.f.][]{Combi2013}. Therefore, more direct evidence is needed to prove/disprove the icy grain halo hypothesis for the case of 209P/LINEAR.

\subsubsection{Nucleus Properties}

As the nucleus is effectively a point source in our data, we reconstruct the nucleus signal by fitting the isolated nucleus signal in Figure~\ref{fig:mdl-mass} with a Gaussian function. This yields the nucleus flux $\mathcal{F_\mathrm{nucleus}}=0.02$~Jy. As the nucleus size has been reliably measured by radar, we derive the corresponding geometric albedo of the nucleus by 

\begin{equation}
A_\lambda(0^\circ) = \left( \frac{r_\mathrm{H}}{1~\mathrm{AU}} \right)^2 \frac{\varDelta^2}{R_\mathrm{N}^2 \Phi(\alpha)} \frac{F_\lambda}{F_{\odot, \lambda}}
\end{equation}

where $\Phi(\alpha)=10^{-0.4\beta\alpha}$ is the phase angle function, with $\alpha$ the phase angle and $\beta=0.035~\mathrm{mag \cdot deg^{-1}}$ the phase slope \citep[e.g.][]{Gehrels1979e}. We yield $A_\lambda(0^\circ)=0.12$ in the $K_\mathrm{s}$ band. This implies a steep spectral slope considering the $R_\mathrm{C}$-band albedo constrained by \citet{Ishiguro2015} that is at the order of 0.05, making the nucleus of 209P/LINEAR similar to D-type asteroids and most Trojan asteroids \citep{Dumas1998a}.

\section{The Meteors}

\subsection{Instrument and Data Acquisition}

The Camelopardalid meteor shower was observed using the Canadian Meteor Orbit Radar (CMOR). CMOR is an interferometric radar array located near London, Canada. The main component of CMOR consists of six stations operated at 29.85~MHz with a pulse repetition frequency of 532~Hz. Meteors are detected along a great circle on the sky plane perpendicular to the radiant vector, when their ionized trails reflect the radar waves sent by the transmitter. Observations are routinely processed by an automatic pipeline to eliminate false detections and calculate trajectory solutions. The details of the CMOR operation can be found in \citet{Jones2005a, Brown2008g} and \citet{Weryk2012}.

\begin{figure}
\includegraphics[width=0.5\textwidth]{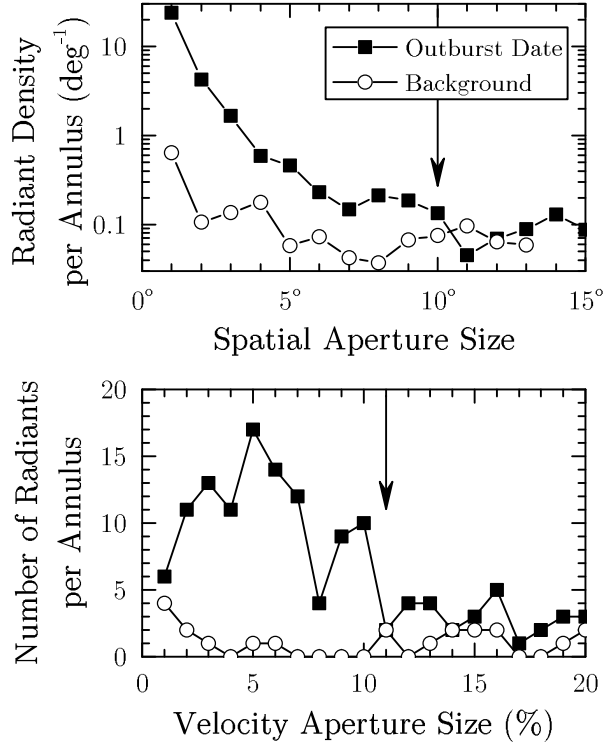}
\caption{Determination of the optimal radiant and velocity apertures. Radiant aperture is centered at $\lambda-\lambda_\odot=38^{\circ}$, $\beta=+57^{\circ}$ in the Sun-centered ecliptic coordinate system, (in-atmosphere) velocity aperture is centered at $v_{\mathrm{m}}=18.8~\mathrm{km \cdot s^{-1}}$. Background values are extracted from non-outburst dates $\pm 2$~days from the outburst date (i.e. 2014 May 22 and 26). The optimal radiant and speed apertures are determined to be $10^{\circ}$ and $11\%$ respectively (marked by arrows). The velocity aperture is determined for the spatial aperture of $10^\circ$.}
\label{fig:met-probe}
\end{figure}

In this study, we focus on multi-station data as it allows for reliable determination of many meteoroid properties. Single-station data (from the main site) is only used for flux calculation. We first prepare our \textit{initial} dataset by extracting Camelopardalid meteors from the processed daily multi-station data, following the procedure described in \citet{Ye2014c}. The aperture (both spatial and velocity) are initially set to the predicted value by \citet{2014MNRAS.437.3283Y} and iterated several times until the optimal values (i.e. includes a maximum number of meteors) are found. A Monte Carlo procedure \citep{Weryk2012} is then used to determine the weighted mean radiant and meteor velocity, which are found to be $\lambda-\lambda_\odot=38^{\circ}$, $\beta=+57^{\circ}$ in the sun-centered coordinate system and with an in-atmosphere velocity $v_\mathrm{m}=18.8~\mathrm{km \cdot s^{-1}}$. The sizes of the spatial and velocity apertures are then found by comparing to the radiant/velocity density profile between the outburst date and the background as determined from ambient meteor activity $\pm 2$~days away from the outburst date. As shown in Figure~\ref{fig:met-probe}, spatial and velocity aperture sizes are determined to be $10^\circ$ and $11\%$ of $v_\mathrm{m}$. A total of 99 Camelopardalid meteors are selected in such manner.

The meteor population studied by CMOR can be broadly classified into \textit{overdense} and \textit{underdense} meteors \citep[e.g.][]{McKinley1961a}. For meteors with similar compositions and properties, overdense meteors are typically associated with larger meteoroids and vice versa. For the case of the Camelopardalid meteor shower, the size cutoff between underdense and overdense meteors is approximately $\beta_\mathrm{rp}=0.0003$ (equivalent to $a_\mathrm{d}=2$~mm assuming $\rho_\mathrm{d}=1000~\mathrm{kg \cdot m^{-3}}$). Compared to the underdense meteors, whose appearance are usually simple, overdense meteors tend to exhibit a complicated and variable appearance, making them sometimes difficult to be identified automatically. Therefore, we retrieved and inspected the raw data $\pm 6$~hr from the predicted peak of the meteor outburst for overdense meteors. A total of 63 Camelopardalid overdense meteors are manually identified in such manner, labeled as the \textit{overdense} dataset. Out of these 63 meteors, 14 of them are also found in the \textit{initial} dataset. We remove these 14 meteors from the \textit{initial} dataset, leaving the other 85 underdense meteors, and label them as the \textit{underdense} dataset. The three datasets are summarized in Table~\ref{tbl:met-dataset-summary}.

\subsection{Results and Analysis of the 2014 Outburst}

\subsubsection{General Characteristics}
\label{sec:met-general}

\begin{figure*}
\includegraphics[width=\textwidth]{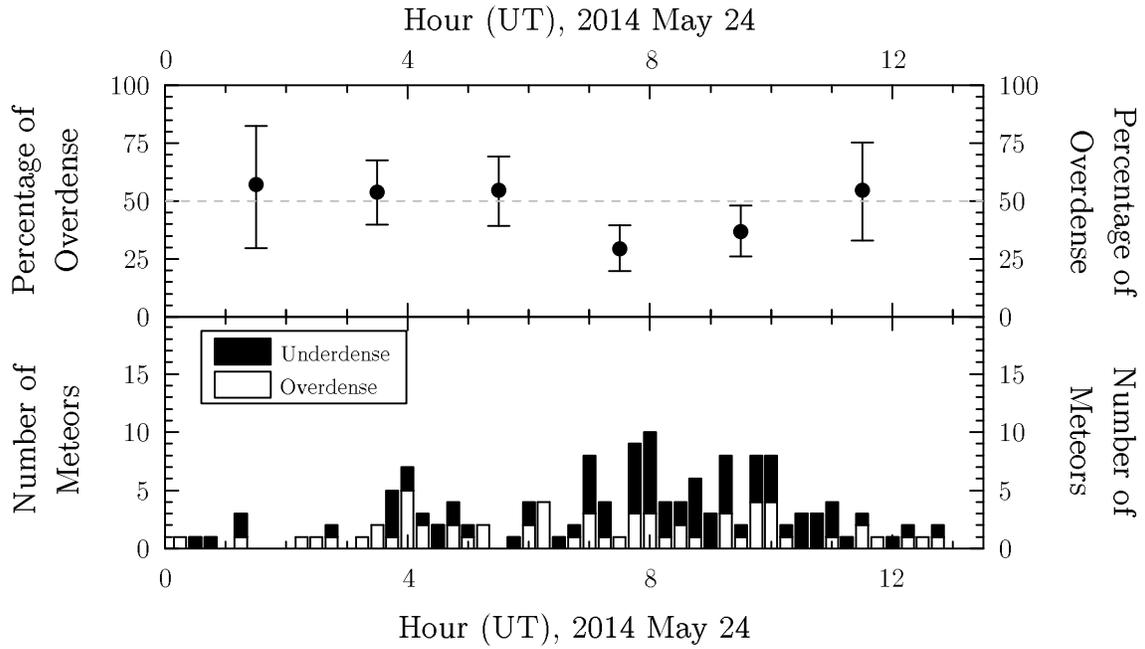}
\caption{Top: Variations of the overdense meteor fraction with Poisson errors, binned in 2~h intervals. A dip (i.e. larger proportion of small meteoroids) is apparent around the peak hour (7--8h UT). Bottom: Raw numbers of overdense and underdense Camelopardalid meteors detected by CMOR, binned in 15~min intervals.}
\label{fig:met-odud}
\end{figure*}

We derive a weighted mean geocentric radiant of $\alpha_\mathrm{G}=124.9^\circ \pm 1.0^\circ$, $\delta_\mathrm{G}=79.2^\circ \pm 0.2^\circ$ (J2000 epoch) and in-atmosphere velocity $v_\mathrm{m}=18.8\pm0.1~\mathrm{km \cdot s^{-1}}$, using the 99 Camelopardalid meteors in the \textit{initial} dataset. This is consistent with the values derived by other studies \citep{Jenniskens2014,Madiedo2014l,Younger2015}. We also note a change in the percentages of overdense and underdense meteors around the peak hour (Figure~\ref{fig:met-odud}), which may reflect the dynamical delivery of meteoroids at different sizes to the Earth's orbit.

\begin{figure}
\includegraphics[width=0.5\textwidth]{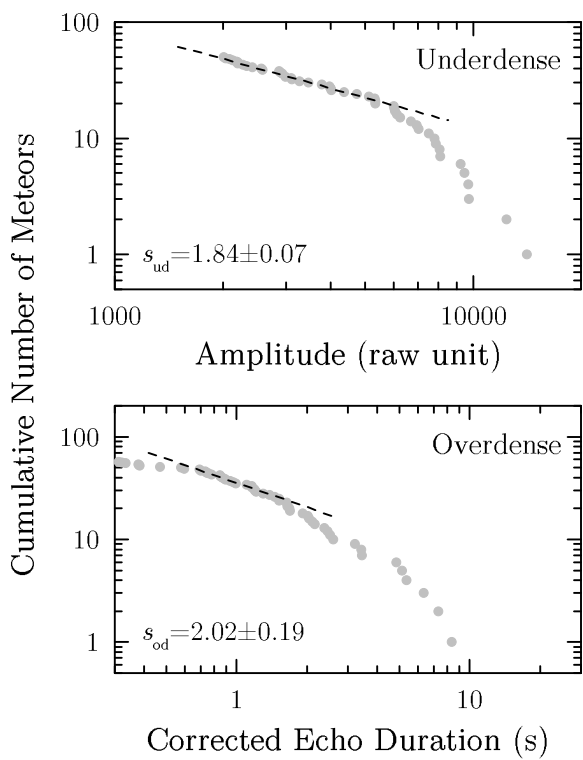}
\caption{Determination of mass indices for the underdense (upper figure) and overdense (lower figure) populations. The mass indices are determined to be $1.84 \pm 0.07$ for unserdense and $2.02 \pm 0.19$ for overdense meteors. The dashed lines show the best fit as determined by the technique developed by (Pokorn\'{y} \& Brown 2015, in prep). The uncertainties are based on the distributions of the posterior probabilities obtained by the MultiNest algorithm \citep{Feroz2013}. The correction of echo duration is described in \citet{Ye2013}.}
\label{fig:met-mi}
\end{figure}

We then derive the mass distribution index $s$ (defined as $\mathrm{d} N \propto m^{-s} \mathrm{d} m$ where $m$ is the mass) for the underdense and overdense population respectively. For underdense meteors, the cumulative amplitude-number relation is typically used to derive the shower mass index \citep[e.g.][]{Blaauw2011a}; for overdense meteors, the cumulative duration-number relation is sometimes used \citep[e.g.][]{McIntosh1968, Ye2014c}. For our underdense sample, we select 50 underdense meteors with echo range within 110--130~km; the range filter is applied to avoid contamination from overdense transition echoes \citep{Blaauw2011a}. For the overdense sample, all 63 meteors in the \textit{overdense} dataset are used. The data and the uncertainty are fitted using the MultiNest algorithm \citep{Feroz2013}, taking account the number statistics of the data. The technique will be described in a separate paper in more detail (Pokorn\'{y} \& Brown, in prep). We find $s_\mathrm{ud}=1.84\pm0.07$ and $s_\mathrm{od}=2.02\pm0.19$ for underdense and overdense meteors respectively (Figure~\ref{fig:met-mi}). This can be related to the size index $q$ by

\begin{equation}
q = 3s-2
\end{equation}

which, for our range of observed $s$, corresponds to $q=3.5$ to 4.1. This agrees with the number derived from cometary observations in \S~\ref{sec:dust_model}.

\begin{figure*}
\includegraphics[width=\textwidth]{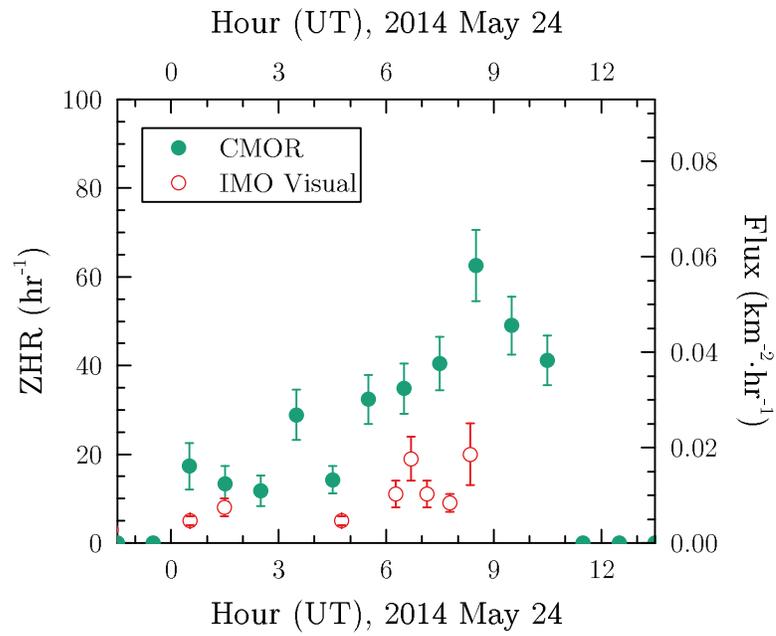}
\caption{The variation of the flux (corrected to a limiting magnitude of +6.5) of the 2014 Camelopardalid meteor outburst as observed by CMOR and IMO visual observers. The CMOR observations are binned in 1~hr intervals. Error bars denoting Poisson errors.}
\label{fig:met-flux}
\end{figure*}

The flux is calculated from the number of meteors detected per unit time divided by the effective collecting area of the radar system, following the procedure described in \citet{Brown1995l}. The calculation of flux does not require a multi-station setup; single-station data is usually sufficient with proper background subtraction. In fact, by using the main-station detections, the statistics can be raised by a factor of $\sim5$. To derive Camelopardalid-only flux, we subtract the raw meteor flux by the background flux following the procedure described in \citet{Ye2013} and \citet{Campbell-Brown2015}. The flux is converted to a Zenith Hourly Rate (ZHR) assuming a single power law size distribution applies to the observed size range \citep{Koschack1990}. The derived CMOR flux is shown in Figure~\ref{fig:met-flux} along with the flux derived from visual observations\footnote{Available at \url{http://www.imo.net/live/cameleopardalids2014/}, retrieved on 2015 April 2.}. Overall, radar and visual observations show agreement in terms of activity timing, with a moderate rise and a steep decline in rates, as well as a main peak around 8h UT, 2014 May 24. We note that the visual profile suffers from small statistics (only $\sim 15$ meteors per bin during the peak, comparing to $\sim 60$ meteors for the radar), and so the two ``peak-lets'' at 6:30 and 8:30 UT are likely artifacts. In both techniques, further refinement of the exact peak time is perhaps not meaningful due to the relatively small statistics of the data. The CMOR flux (corrected to a limiting magnitude of +6.5) is about half an order of magnitude higher than the visual flux, seemingly indicating an overabundance of faint meteors and a break in the power law somewhere beyond the naked-eye limit.

\subsubsection{Meteoroid Properties}

\begin{figure*}
\includegraphics[width=\textwidth]{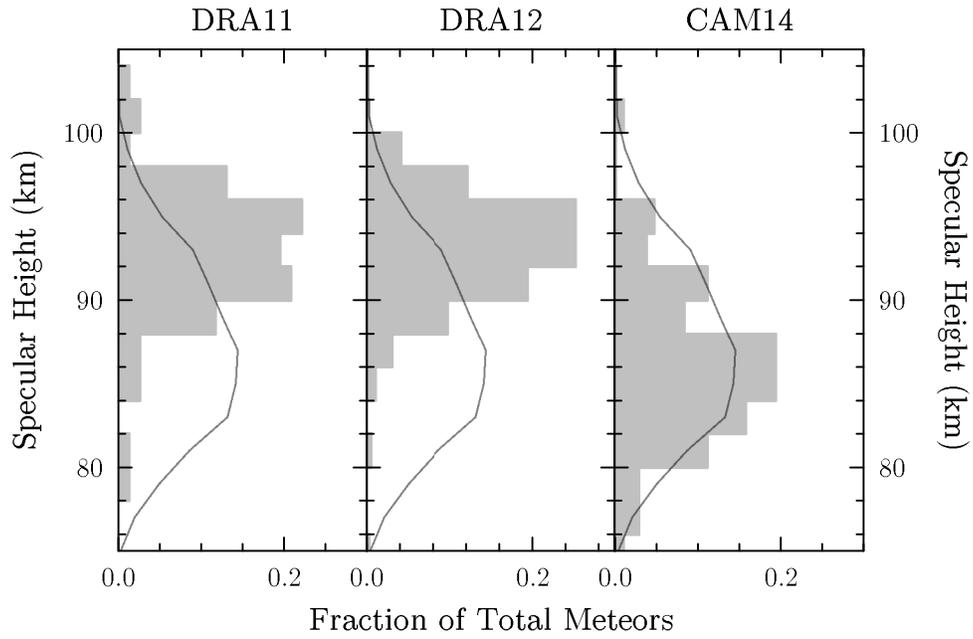}
\caption{Specular height distribution of the underdense meteor echoes observed by CMOR for the 2011/12 Draconid outbursts (denoted as DRA11 and DRA12) and 2014 Camelopardalid outburst (denoted as CAM14), plotted as shaded bars. Specular height distribution of sporadic meteors (generated using all meteors detected by CMOR with $v_\mathrm{m}$ within 5\% from $20~\mathrm{km \cdot s^{-1}}$) is shown as line.}
\label{fig:met-ht}
\end{figure*}

The Camelopardalids have almost identical entry speeds and geometry (with respect to CMOR) as another JFC shower, the October Draconids, which was observed by CMOR during its 2011 and 2012 outbursts \citep{Ye2013, Ye2014c}. This coincidence allows us to directly compare the main characteristics of these two showers independent of instrumental effects or entry speed corrections. A distinct difference between the two showers is in the specular height distribution of the meteors: the Draconids appear 5--10~km higher than the Camelopardalids as observed by CMOR (Figure~\ref{fig:met-ht}). It has long been thought that the exceptional ablation height of the Draconids is the direct consequence of their extreme fragility \citep[e.g.][]{Borovivcka2007}. Hence, a simple interpretation of the observed height distribution of the two showers is that the Camelopardalid meteoroids are less fragile relative to the Draconids. As the outbursts from the two showers originated from cometary ejecta with young ejection ages (less than a few hundred years), the difference in space weathering is not significant; the observations seem to suggest that the surface material properties of the two parent comets are different.

We compare our result to the results derived from other Camelopardalid studies. \citet{Younger2015}, who also observed the 2014 Camelopardalid outburst with a meteor radar, reported that the Camelopardalid meteoroids were less fragile than sporadic meteoroids, a finding that is not apparent in our Figure~\ref{fig:met-ht} due to our aggressive binning to enhance the statistics; but Younger et al.'s finding is at least qualitatively consistent with our finding that the Camelopardalid meteoroids being less fragile relative to the Draconids. Conversely, optical observations by \citet{Jenniskens2014} and \citet{Madiedo2014l} show that the Camelopardalid meteoroids are very fragile and are consistent with fluffy aggregates like the Draconids. However, we note that (1) optical observations are sampling meteoroids of a larger size range (close to centimeter-sized, while radar observations are sampling millimeter-sized meteoroids); and (2) \citet{Jenniskens2014} and \citet{Madiedo2014l}'s observed meteors were recorded in a wider time span than the radar (on the order of 1~d vs. a few hours). Meteors detected away from the predicted peak mainly consist of background meteoroids that are part of older, disrupted trails. Hence, the optical meteors, whose properties seem very different than the radar meteors, may represent Camelopardalid meteoroids at different sizes and ages.

\subsection{Camelopardalid Activity in Other Years}

We conduct a search in the CMOR database for any undetected Camelopardalid activity in previous years, using the 3-dimensional wavelet analysis technique \citep[e.g.][]{Brown2010q,Bruzzone2015} to compute the wavelet coefficient at the location of the Camelopardalid radiant. The time window is restricted to one week around the nodal passage of 209P, namely in the solar longitude range $\lambda_\odot=60^\circ-66^\circ$. CMOR has been fully operational since 2002, but data in 2002, 2006, 2009 and 2010 are severely (off-line periods more than 24 hours) interrupted by instrumental issues; hence we only inspect years with complete data for possible Camelopardalid activity.

\begin{figure}
\includegraphics[width=0.5\textwidth]{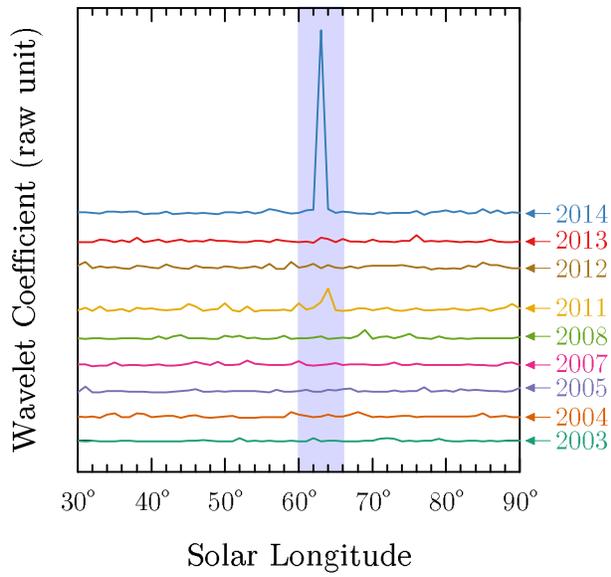}
\caption{Variation of the relative wavelet coefficient at $\lambda-\lambda_\odot=38^{\circ}$, $\beta=+57^{\circ}$ and $v=20~\mathrm{km \cdot s^{-1}}$ within $\lambda_\odot=30^\circ-90^\circ$ in 2003--2014 (except 2006, 2009 and 2010). The expected Camelopardalid activity period is shaded. Activity is noticeable only in 2011 and 2014.}
\label{fig:met-wc}
\end{figure}

\begin{figure}
\includegraphics[width=0.5\textwidth]{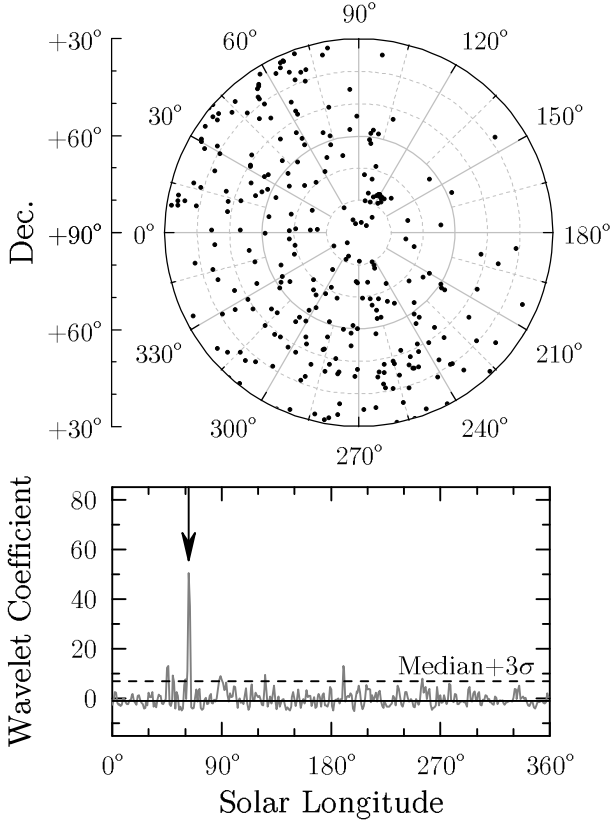}
\caption{Upper figure: the raw radiant map of all meteor echoes detected by CMOR on 2011 May 25, corresponding to solar longitude $\lambda_\odot=63^\circ$. Angular axis represents R.A. and the radial axis represents Declination, both in geocentric coordinates in J2000 coordinates. Radiants are plotted as black dots. The Camelopardalid activity is clearly visible near $\alpha_\mathrm{G}=120^\circ, \delta_\mathrm{G}=+80^\circ$. Lower figure: variation of the relative wavelet coefficient at $\lambda-\lambda_\odot=38^{\circ}$, $\beta=+57^{\circ}$ and $v=20~\mathrm{km \cdot s^{-1}}$ in 2011, with the Camelopardalid activity marked by an arrow. Solid and dashed lines are median and $3\sigma$ above median, respectively.}
\label{fig:met-2011-raw}
\end{figure}

We find distinct activity in 2011, while the activity in other years, if any, was too weak to be reliably separated from the background (Figure~\ref{fig:met-wc}). The 2011 outburst is even noticeable on the raw, unprocessed radiant map (Figure~\ref{fig:met-2011-raw}), albeit much weaker than the 2014 outburst. We were able to extract 15 meteors for the 2011 outburst, which yields a weighted radiant of $\alpha_\mathrm{G}=119.5^\circ \pm 2.1^\circ$, $\delta_\mathrm{G}=77.2^\circ \pm 0.3^\circ$ (J2000 epoch) and in-atmosphere velocity $v_\mathrm{m}=19.3\pm0.3~\mathrm{km \cdot s^{-1}}$. We find no obvious peak of activity, but the core of the activity falls between 2011 May 25 at 6--11~h UT ($\lambda_\odot=63.6^\circ$). The 2011 activity was not high enough to derive a statistically meaningful flux, but we estimate the 2011 flux to be about an order of magnitude lower than the 2014 flux, since the number of raw echoes is roughly $1/10$ of that of 2014\footnote{The change of radar collecting area in different years is negligible thanks to the high declination of the Camelopardalid radiant.}. By following the same technique described in \S~\ref{sec:met-general}, we derive a $1\sigma$ upper limit of the flux to be $\lesssim0.01~\mathrm{km^{-2} \cdot hr^{-1}}$ for other years. 

\subsection{Modeling the Dust (II)}
\label{sec:met-model}

The dust model derived from cometary observations has placed some useful constraints on the physical properties of the Camelopardalid meteoroids. In this section, we explore the contribution of young meteoroid trails (defined as trails formed within $\sim 50$~orbital revolutions) to the observed meteor activity using numerical techniques. Older dust trails have experienced more perturbations from the major planets and are too disrupted to model. The simulation procedure is essentially the same as that in \S~\ref{sec:dust_model}, apart from extending the integration time several hundred years backward. To address possible meteor activity, we select a subset of Earth-approaching meteoroids following the method discussed by \citet{Brown1998d} and \citet{Vaubaillon2005a}:

\begin{equation}
 \Delta X = v_\mathrm{rel} \times \Delta T
\end{equation}

where $v_\mathrm{rel}\approx17~\mathrm{km \cdot s^{-1}}$ is the relative velocity between the meteoroid and the Earth, $\Delta T$ is the characteristic duration of the meteor shower which we take as $\Delta T=1$~d. These yield $\Delta X=0.01$~AU. The simulated meteoroid is included in the subset when its Minimum Orbit Intersection Distance (MOID) to the Earth's orbit, calculated with the subroutine developed by \citet{Gronchi2005a}, is smaller than $\Delta X$.

We use the dust model derived from our cometary observations for the ejection of meteoroids. For comparison, the traditional \citet{Crifo1997a} model (denoted as the C\&R model hereafter) is also used in a parallel simulation. The start of the integration is set to 50 orbits ago (or about 1750 A.D.). We first integrate 209P/LINEAR back to the year of 1750, and then integrate it forward with meteoroids released at each perihelion passage when the parent has $r_\mathrm{H}<1.4$~AU, the heliocentric limit of cometary activity as indicated by cometary observations. When the simulation is finished, we examine the encounters of all meteoroid trails in the years that CMOR was operational.

\begin{figure*}
\includegraphics[width=\textwidth]{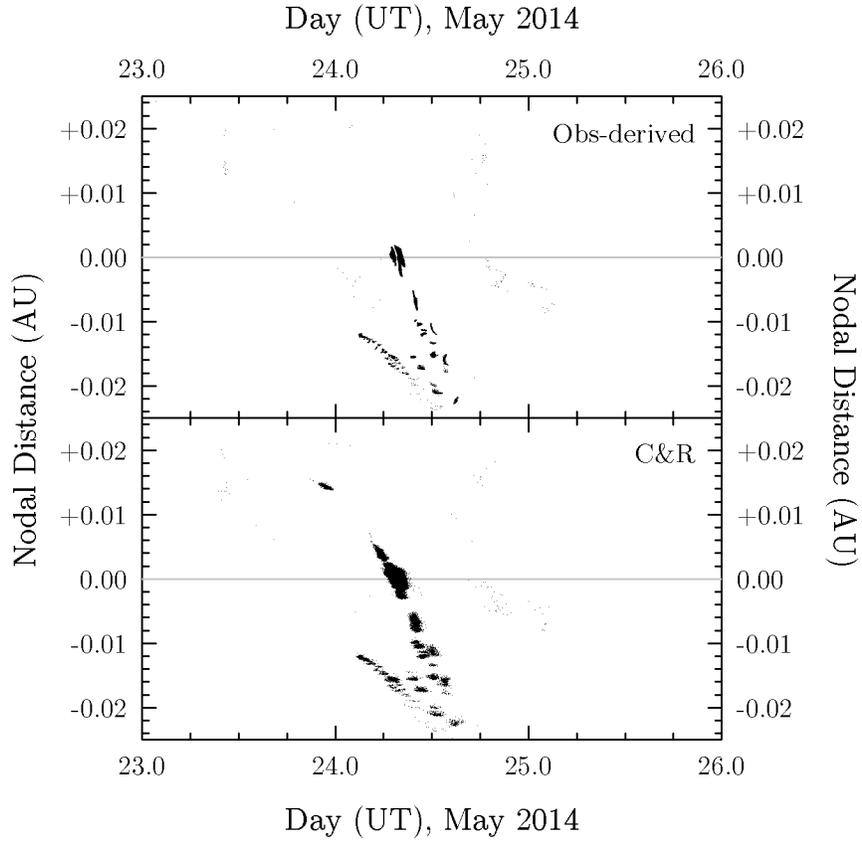}
\caption{Nodal footprint of the 1750--2000 trails around 2014 May 24, using the ejection model derived from comet observations (upper figure) and the \citet{Crifo1997a} ejection model (lower figure).}
\label{fig:met2014sim}
\end{figure*}

\begin{figure*}
\includegraphics[width=\textwidth]{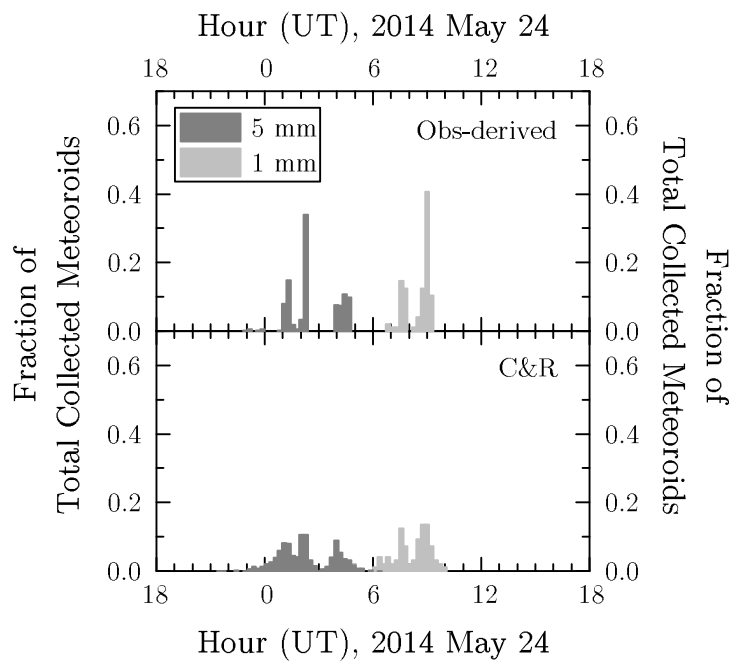}
\caption{The arrival distribution of large, overdense-like ($a_\mathrm{d}=5~\mathrm{mm}$ or $\beta_\mathrm{rp}=0.0001$) and small, underdense-like ($a_\mathrm{d}=1~\mathrm{mm}$ or $\beta_\mathrm{rp}=0.0005$) meteoroids from observation-derived (upper figure) and the \citet{Crifo1997a} ejection models (lower figure) for the 2014 Camelopardalid meteor outburst. It is apparent that larger meteoroids arrived earlier than smaller meteoroids, consistent with CMOR observations.}
\label{fig:met2014sim-size}
\end{figure*}

The results from both ejection models are largely identical, making it difficult to distinguish the better ejection model using observations. This emphasizes that the evolution of older trails is predominantly controlled by planetary perturbations rather than ejection speed. The 2014 encounter is easily identifiable thanks to the high density of the corresponding trail (Figure~\ref{fig:met2014sim}), with the simulation agreeing with the observations. We also note that our simulation predicts the Earth would first encounter larger meteoroids (Figure~\ref{fig:met2014sim-size}), a result consistent with CMOR observation of early overdense meteors noted in \S~\ref{sec:met-general}.

The flux of meteoroids can be estimated by relating the number of meteoroids in Earth's vicinity to the dust production rate of the comet. From the analysis in \S~\ref{sec:near_nucleus}, we estimate the current dust production rate of 209P/LINEAR is of the order of $10^6$~kg, or $N\sim10^{14}$ meteoroids per orbit (taking $\bar{a}\sim10^{-4}$~m as found previously). From the meteoroid stream simulation, we find $\sim 1\%$ of the meteoroids released between 1750--2014 are delivered to the Earth's vicinity during the 2014 encounter, corresponding to a flux of $\mathcal{F} \sim 1\% \times N \times \Delta X^{-2} \times \Delta T^{-1}=0.01 \times 10^{14} \times (0.01~\mathrm{AU})^{-2} \times (1~\mathrm{d})^{-1}=0.02~\mathrm{km^{-2} \cdot hr^{-1}}$, comparable to the flux determined from visual and radar meteor observations. This implies that 209P/LINEAR was not substantially much more active in the past several centuries, an idea also supported by the apparent lack of annual activity of the Camelopardalid meteor shower.

\begin{figure*}
\includegraphics[width=\textwidth]{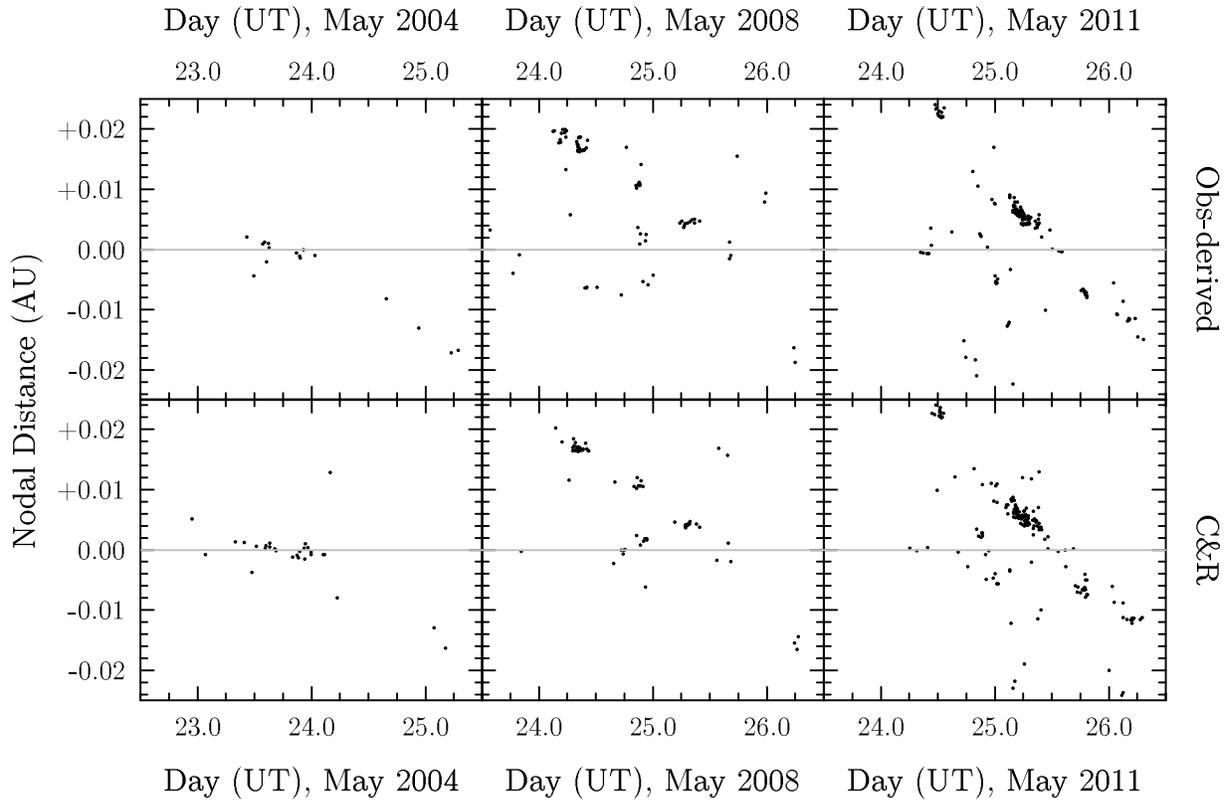}
\caption{Nodal footprint of the 1750--2000 trails around 2004 May 24, 2008 May 25 and 2011 May 25, using the ejection model derived from comet observations (upper row) and the \citet{Crifo1997a} ejection model (lower row). The scale of meteoroid number is identical to that of Figure~\ref{fig:met2014sim}, but for clarity the meteoroids in this figure are marked with larger symbols.}
\label{fig:metpre2014sim}
\end{figure*}

Additionally, we find predicted encounters in 2004, 2008 and 2011 from our simulations (Figure~\ref{fig:metpre2014sim}). The 2004 and 2008 encounters are predicted to be about an order of magnitude weaker than the 2011 encounter, thus we expect this activity to be buried in the sporadic background. The 2011 case is interesting as the parent was near aphelion at the time of the meteor outburst. Both the C\&R model and our ejection model derived from the cometary observations only indicate encounters with a few extremely weak trails formed between 1763--1768 in 2011. The calculated peak time and width (both ejection models suggest peak times of 2011 May 25 $\sim$5:40 and 9:00 UT for the 1763- and 1768-trail, with full-width-half-maximum of $\sim 8$~hr) is consistent with CMOR observations. However, the flux predicted by the model is by a factor of 100 lower than what was observed, potentially hinting at a significant but transient increase of activity of 209P/LINEAR around those epochs. The same 1763- and 1768-trail also contribute to the 2014 meteor event; however, the overlapping peak time between trails (mostly $<1$~hr apart) makes it difficult to distinguish activity from individual trails in the observations.

\section{Discussion}

\subsection{The Dynamical Evolution of 209P/LINEAR}

Recent work by \citet{Fernandez2014j} revealed a set of unique members among the JFCs that reside in highly stable ($>10^4$~yr) orbits, including 209P/LINEAR. We extend their work for the case of 209P/LINEAR by generating 1000~clones of 209P/LINEAR using the orbital covariance matrix provided in JPL~130, and integrate all of them $10^5$~yr backwards. The integration is performed with MERCURY6 using the Bulirsch-Stoer integrator \citep{Bulirsch1972,Stoer1972}.

\begin{figure*}
\includegraphics[width=\textwidth]{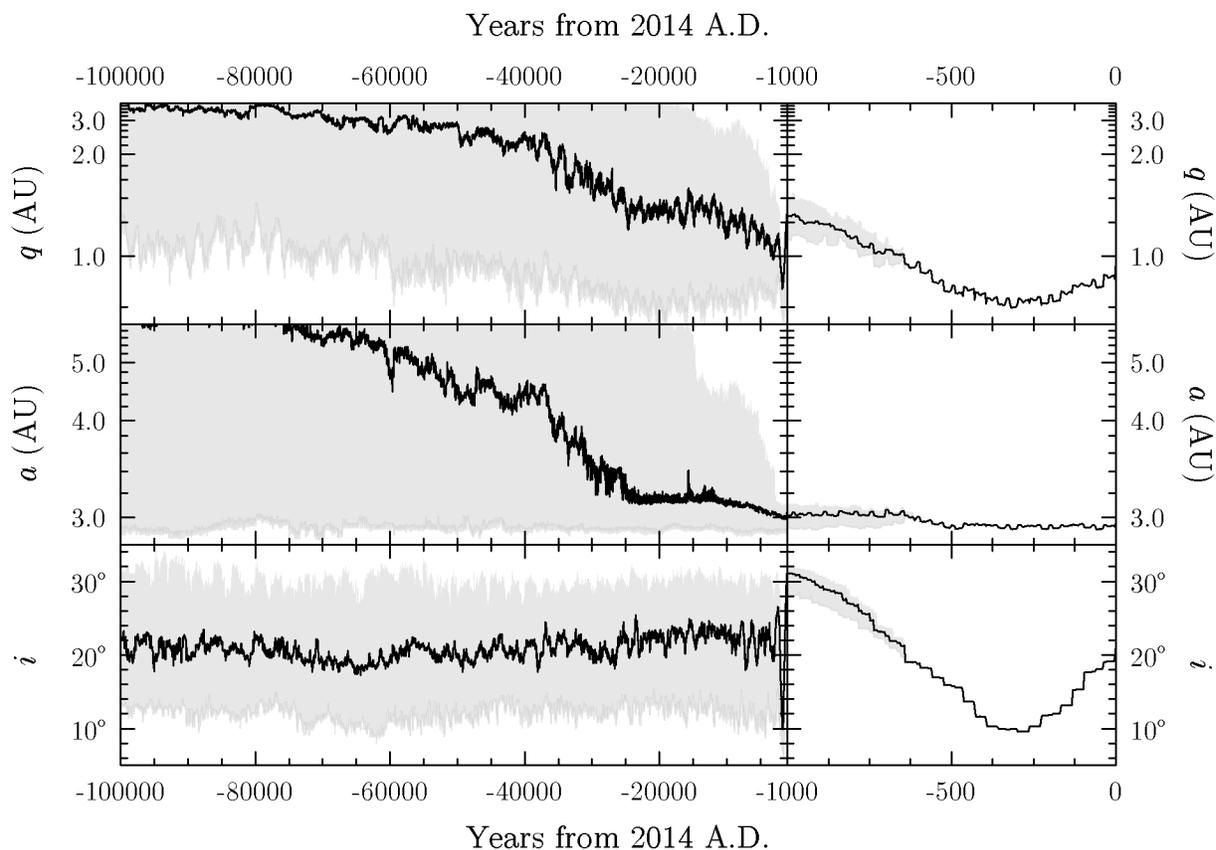}
\caption{Dynamical evolution of 1000 clones of 209P/LINEAR in a time interval of $10^5$~yr with a zoomed section for within 1000~yr. The median (black line) and $\pm1\sigma$ region (shaded area) is shown. A highly stable section is seen up to $3\times10^4$~years, of which the core of the clones remain in near-Earth region and $95\%$ of the clones remain in bounded orbits.}
\label{fig:img-comet-dyn}
\end{figure*}

As shown in Figure~\ref{fig:img-comet-dyn}, the core of the clones remain in Earth's vicinity for $\sim 10^4$~years, much longer than the typical physical lifetime for similar-sized JFCs in the near-Earth region \citep[e.g.][]{2009Icar..203..140D}. In addition, we note the core of the clones is extremely compact for more than 100 orbits ($1\sigma$ width in semimajor axis $\sim0.0002$~AU), until an extreme close approach to the Earth ($d_\mathrm{min}\approx0.006_{-0.005}^{+0.010}$~AU) around 1400 Mar 12 (on Julian calendar) scatters the clones. The miss distance of this approach to the Earth is smaller than the recorded close approach by Lexell's Comet in 1770 \citep[0.015~AU;][]{Kronk2008} and prompted us to look at medieval astronomical records for possible sightings, without success. If the activity level of 209P/LINEAR in the 15th century is comparable to what it is now, the comet would have been $+7$~mag during its approach in 1400, below the naked-eye limit of medieval astronomers; however, any significant (by several magnitudes) increase in activity could have been noticeable. The lack of possible sightings for 209P/LINEAR's close approach in 1400 suggests that the comet was not substantially more active $\sim 100$~orbits ago.

\begin{figure*}
\includegraphics[width=\textwidth]{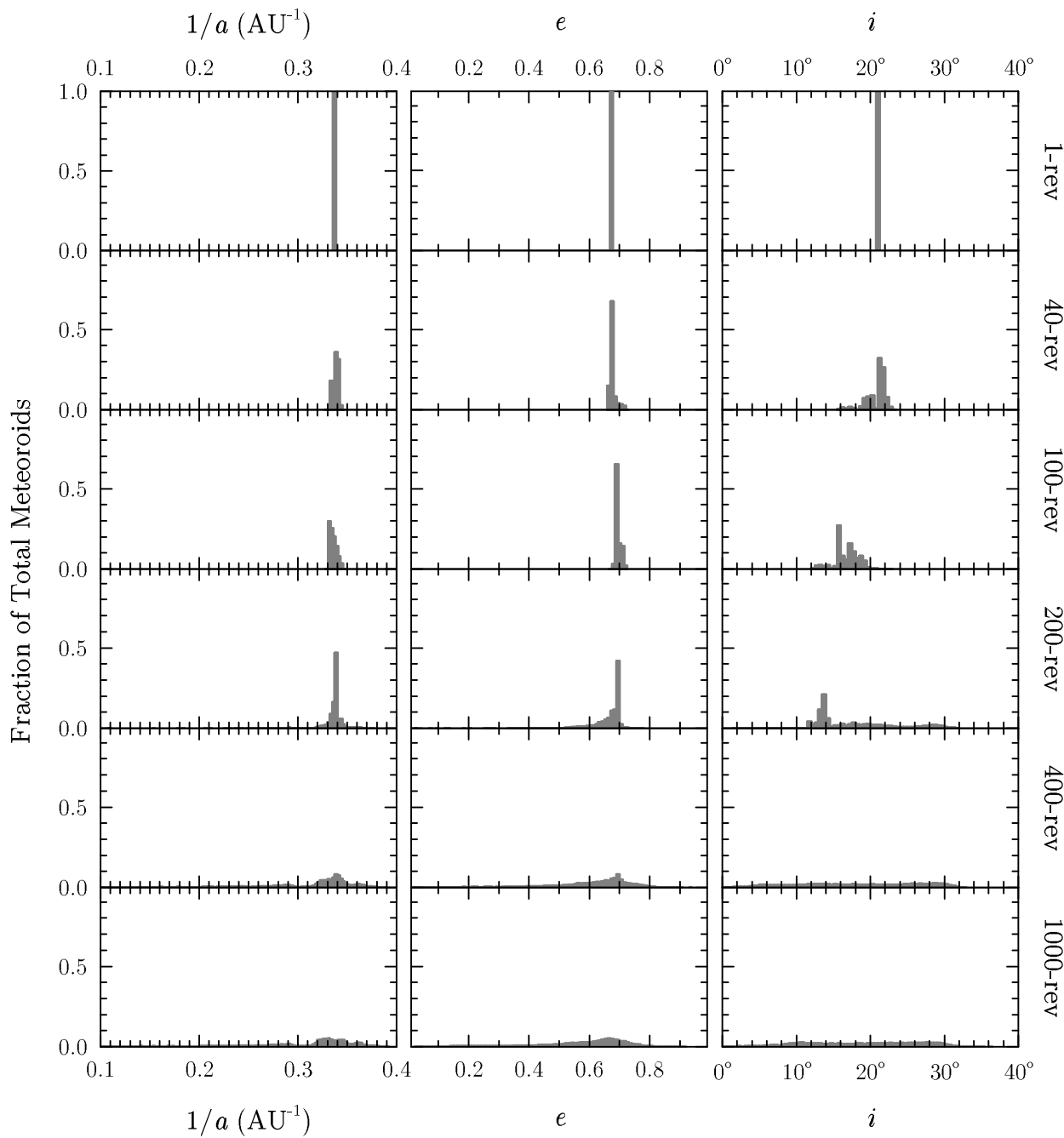}
\caption{Evolution of secular orbital elements of meteoroids of different ages: 1-rev (meteoroids released 5~yr ago), 40-rev (released 200~yr ago), 100-rev (released 500~yr ago), 200-rev (released 1000~yr ago), 400-rev (released 2000~yr ago) and 1000-rev (released 5000~yr ago). The meteoroid ejection model is based on comet observations, but the result is insensitive to the choice of ejection model, as the evolution of meteoroid stream is predominantly controlled by planetary perturbations over the investigated time scale. It can be seen that the dispersion time scale of the Camelopardalid meteoroid stream is at the order of 1000~yr (200-rev).}
\label{fig:met-pre2014}
\end{figure*}

Since 209P/LINEAR is in a stable orbit, the associated meteoroid stream may also possess a set of orbits that are more stable than other JFC streams. To quantify the dispersion process of the Camelopardalid meteoroid stream, we adopt the same integration procedure as described in \S~\ref{sec:met-model} and examine the evolution of meteoroid trails released between 1-revolution (5~yr) and 1000-revolution (5000~yr), shown as Figure~\ref{fig:met-pre2014}. It can be seen that the narrow stream structure is maintained for trails that formed as far as $\sim 1000$ to 2000~yr ago, which is a few times longer than other JFC streams such as the $\pi$-Puppid meteoroid stream \citep[e.g.][]{1997A&A...324..770C}. We also note that the meteoroid stream evolves differently than the parent. The degree of the difference increases as the age of the stream increase. For example, the current radiant of the core of 200-rev meteoroids (i.e. meteoroids released at about 1000~A.D.) would be at $\alpha_\mathrm{G}=120^\circ$, $\delta_\mathrm{G}=+60^\circ$, encountered at $\lambda_\odot=70^\circ$ (approximately June 1). There is no established meteor activity related to this hypothetical radiant, although a few other possible annual showers have been associated with 209P/LINEAR \citep[e.g.][]{Rudawska2014g,vSegon2014}.

\subsection{Nature of 209P/LINEAR and Comparison with Other Low Activity Comets}

Following our analysis, it seems evident that 209P/LINEAR has been mostly weakly active for the last few hundred orbits, while it might have been in a near-Earth JFC orbit on the time scale of $\sim 10^4$~yr. This is compatible with the idea of 209P/LINEAR as an aging comet exhausting its remaining near surface volatiles as derived from the classical interpretation of cometary evolution. It is perhaps not possible to know how long the comet has stayed in the inner solar system; however, we note that the gradual decrease of the perihelion over the course of few thousand years (as indicated in Figure~\ref{fig:img-comet-dyn}) may provide a prolonged favorable environment for weak cometary activity, as the sub-surface volatiles underneath the dust mantles can be (re-)activated by the gentle decrease of the perihelion distance \citep{Rickman1990e}.

What does 209P/LINEAR tell us about other low activity comets? In the following we briefly discuss three other Earth-approaching comets (i.e. those that may generate meteor showers) listed in Table~\ref{tbl:wc} and compare them to 209P/LINEAR. The other five comets in the list do not generate meteor showers, making it difficult to address their physical history in a manner similar to 209P/LINEAR.

\paragraph{252P/LINEAR}

Little is known about this newly discovered comet at the moment, except that numerical simulation indicate a recent ($<100$~orbits) entry to the inner solar system \citep[][see also \url{http://www.astronomia.edu.uy/Criterion/Comets/Dynamics/table_num.html}, retrieved 2015 May 17]{Tancredi2014b}, implying a different origin and evolution compared to 209P/LINEAR. Considering its young dynamical age in the inner solar system, the low activity of 252P/LINEAR may reflect a relative lack of volatiles at the time of formation of the nucleus.

\paragraph{289P/Blanpain}

289P/Blanpain is the only low activity comet in the list that is associated unambiguously with annual meteor activity \citep{Jenniskens2008l}. The comet itself was lost for some 200~yr after its initial discovery in 1819 (and had been referred as D/1819 W1), until being re-discovered as a faint asteroidal body 2003 WY25 in 2005 \citep{Foglia2005}. Multiple clues suggest 2003 WY25 is the remnant of the original 289P/Blanpain following a catastrophic fragmentation event \citep[e.g.][]{Jenniskens2005e,Jewitt2006g}. Hence, the low activity nature of 289P/Blanpain may have a completely different origin than that of 209P/LINEAR.

\paragraph{300P/Catalina}

300P/Catalina (known as 2005 JQ5 in some early literatures) is interesting, as it is the only other comet in our list that is concurrently classified as a stable JFC by \citet{Fernandez2014j}. It has not been associated with any established annual meteor shower, although a few possible linkages have been suggested \citep[e.g.][]{Rudawska2014g}. Radar observations by \citet{Harmon2006e} revealed a rough surface similar to 209P/LINEAR; however, the presence of cm-sized dust around the nucleus of 300P/Catalina, which is absent for 209P/LINEAR \citep{Howell2014v}, seems to indicate stronger outgassing activity of 300P/Catalina compared to 209P/LINEAR at the present time. It may be possible that 300P/Catalina is at an earlier stage of dormancy compared to 209P/LINEAR.

\section{Conclusions and Summary}

The low activity comet, 209P/LINEAR, may indeed be an aging comet that is quietly exhausting its last bit of near surface volatiles. This idea is supported by the convergence of several different lines of evidence: dust modeling of cometary images that revealed a presently weakly active comet, analysis and modeling of meteor observations that revealed a low dust production over the past few hundred orbits, numerical analysis of the dynamical evolution of the comet that suggested a stable orbit in the inner solar system over a time scale of $10^4$~yr.

The main findings of this paper are:

\begin{enumerate}
 \item The best-fit dust model to the cometary images involves a low ejection speed ($1/10$ of moderately active comets) and large dust grains ($\bar{a}_\mathrm{d}=10^{-4}$~m). The dust production rate of the comet at 19~d after perihelion is $0.4~\mathrm{kg \cdot s^{-1}}$, a remarkably small number.
 \item The coma region appears to be inconsistent with the steady-flow model. The general characteristics of this region is compatible with the icy grain halo theory, a theory that is known to be only applicable to active long period comets and hyperactive Jupiter-family comets. More conclusive evidence is needed to establish or disprove this hypothesis.
 \item By applying a coma subtraction technique, the nucleus signal is separated from the coma, yielding a geometric albedo $A_\lambda(0^\circ)=0.12$ appropriated to $K_\mathrm{s}$ band. Coupling with optical measurements at visible band, this indicates a reddish spectrum of the nucleus of 209P/LINEAR similar to that of D-type asteroids and most Trojans.
 \item Radar observations by CMOR show the peak of 2014 Camelopardalid meteor outburst around 2014 May 24 at 8~h UT. From CMOR observations, we derive a mean radiant of $\alpha_\mathrm{G}=124.9^\circ \pm 1.0^\circ$, $\delta_\mathrm{G}=79.2^\circ \pm 0.2^\circ$ (J2000 epoch), mean in-atmosphere velocity $v_\mathrm{m}=18.8\pm0.1~\mathrm{km \cdot s^{-1}}$, and a peak flux of $0.06~\mathrm{km^{-2} \cdot hr^{-1}}$, consistent with visual, optical and other radar observations. Numerical simulation confirms that the outburst originated from the dust trails formed in the 18--20th century, a time that the parent was perhaps not much more active. The mass distribution index of the meteors, $s=1.8$~to~$2.0$, agrees the size index $q=3.8$ derived from the modeling of the cometary images.
 \item A direct comparison to the Draconids, a meteor shower with almost identical entry speed that was also observed with CMOR, shows that a distinctly different height distribution between the Camelopardalids and Draconids: the Camelopardalids tend to appear $\sim10\%$ lower than the Draconids. This is likely due to the Camelopardalids being less fragile relative to the Draconids, the latter of which have long been known as extremely fragile meteoroids. This agrees with other radar measurements but differs from optical measurements, which support highly fragile meteoroids. As optical observations are sampling meteoroids at larger sizes and wider arrival times, the difference in meteoroid properties derived from different techniques may be due to sampled meteoroids of different sizes and ages.
 \item We examine CMOR data from 2003 onwards (except 2006, 2009 and 2010) and find a previously unnoticed Camelopardalid outburst in 2011. The activity peaks around 2011 May 25 between 6--11~h UT, with a peak flux of the order of $0.005~\mathrm{km^{-2} \cdot hr^{-1}}$. Numerical simulations suggest the dust trail encountered in 2011 was formed in 1763--1768, however the predicted flux seems to be by a factor of 100 smaller than what was observed. This may indicate some temporary increase in activity of 209P/LINEAR around those times.
 \item Numerical integration indicates 209P/LINEAR may have resided in a stable near-Earth JFC orbit for $\sim 10^4$~yr. The dispersion time scale for the Camelopardalid stream is about 1000--2000~yr, which is a few times longer than JFC streams such as the $\pi$-Puppids. The lack of significant annual activity of the Camelopardalid shower may serve as a strong evidence of the low activity of 209P/LINEAR over the past several hundred orbits. 
 \item We compare 209P/LINEAR to three other low activity comets that are associated with known or hypothetical meteor showers: 252P/LINEAR (associated with a hypothetical meteor shower in the constellation of Lepus), 289P/Blanpain (associated with the Phoenicid meteor shower), and 300P/Catalina (associated with a few possible meteor showers, such as the June $\epsilon$-Ophiuchids). A diversity is seen: the low activity of 252P/LINEAR may be congenital; that of 289P/Blanpain may be due to catastrophic fragmentation. 300P/Catalina shares many similar physical and dynamical characteristics with 209P/LINEAR; but the presence of cm-sized meteoroids around the nucleus may indicate a stronger outgassing activity of 300P/Catalina compared to 209P/LINEAR at the moment.
\end{enumerate}

\acknowledgments

We thank Jian-Yang Li and Iwan Williams for their thorough comments, and Pauline Barmby, Sebasti\'{a}n Bruzzone, Michael Combi, Julio Fern\'{a}ndez, Peter Jenniskens, David Jewitt, Song Huang, J\'{e}r\'{e}mie Vaubaillon and Bin Yang for discussions. We also thank Zbigniew Krzeminski, Jason Gill, Robert Weryk and Daniel Wong for helping with CMOR operations. The work is based on observations obtained at the Gemini Observatory, acquired through the Gemini Science Archive and processed using the Gemini IRAF package, which is operated by the Association of Universities for Research in Astronomy, Inc., under a cooperative agreement with the NSF on behalf of the Gemini partnership: the National Science Foundation (United States), the National Research Council (Canada), CONICYT (Chile), the Australian Research Council (Australia), Minist\'{e}rio da Ci\^{e}ncia, Tecnologia e Inova\c{c}\~{a}o (Brazil) and Ministerio de Ciencia, Tecnolog\'{i}a e Innovaci\'{o}n Productiva (Argentina). Part of the numerical simulations were conducted using the facilities of the Shared Hierarchical Academic Research Computing Network (SHARCNET:www.sharcnet.ca) and Compute/Calcul Canada. Funding support from the NASA Meteoroid Environment Office (cooperative agreement NNX11AB76A) for CMOR operations is gratefully acknowledged.

Additionally, we thank all visual meteor observers for their reports: Tomasz Adam, Jose Alvarellos, Orlando Ben\'{\i}tez S\'anchez, Jens Briesemeister, Mark Davis, Michel Deconinck, Peter Detterline, Jose Vicente D\'{i}­az Mart\'{i}­nez, Audrius Dubietis, Shy Halatzi, Carl Hergenrother, Glenn Hughes, Richard Huziak, Jens Lacorne, Michael Linnolt, Boris Majic, Roman Makhnenko, Aleksandar Matic, Bruce Mccurdy, Jaroslav Merc, Vasilis Metallinos, Milen Minev, Arash Nabizadeh Haghighi, Michael Nolle, Ella Ratz, Kai Schultze, Miguel Angel Serra Martin, Wesley Stone, Richard Taibi, Tamara Tchenak, Sonal Thorve, Koen Miskotte, Tomasz Lenart, Pierre Martin, Alexsandr Morozov, Maciek Myszkiewicz, Salvador Aguirre, Branislav Savic, Javor Kac, Pierre Bader, William Watson, Alexandr Maidik, Anna Levina, Karoly Jonas, Josep M Trigo-Rodriguez, Daniel Verde Van Ouytsel, and Quan-Zhi Ye.


\clearpage

\clearpage

\clearpage

\clearpage

\clearpage

\clearpage

\clearpage

\clearpage

\clearpage

\clearpage

\clearpage

\clearpage

\clearpage

\clearpage

\clearpage

\clearpage

\clearpage

\begin{table*}
\centering
\caption{A list of low activity comets according to the definition given in \S1.}
\begin{tabular}{cccl}
\hline
Comet & $M_1$ & $R_\mathrm{N}$ & Assoc. meteor shower \\
 & & (km) & \\
\hline
10P/Tempel 2 & 13.2 & 10.6\tablenotemark{a} & - \\
28P/Neujmin 1 & 11.5 & 21.4\tablenotemark{a} & - \\
102P/Shoemaker 1 & 15.7 & 1.6\tablenotemark{b} & - \\
184P/Lovas 2 & 14.4 & 6.2\tablenotemark{b} & - \\
209P/LINEAR & 16.9 & 2.7\tablenotemark{c} & Camelopardalids \\
252P/LINEAR & 18.6 & 0.5\tablenotemark{d} & Predicted, not yet observed\tablenotemark{g} \\
289P/Blanpain & 22.9 & 0.32\tablenotemark{e} & Phoenicids \\
300P/Catalina & 18.3 & 1.4\tablenotemark{f} & June $\epsilon$-Ophiuchids (?) \\
C/2001 OG108 (LONEOS) & 13.1 & 13.6\tablenotemark{a} & - \\
\hline
\end{tabular}
\tablenotetext{a}{\citet{Lamy2004}.}
\tablenotetext{b}{\citet{Scotti1994p}.}
\tablenotetext{c}{\citet{Howell2014v}.}
\tablenotetext{d}{Drahus (2015, personal communication).}
\tablenotetext{e}{\citet{Jewitt2006g}.}
\tablenotetext{f}{\citet{Harmon2006e}.}
\tablenotetext{g}{Unpublished data from Maslov (\url{http://feraj.narod.ru/Radiants/Predictions/252p-ids2016eng.html}, retrieved 2015 May 2).}
\label{tbl:wc}
\end{table*}

\clearpage

\begin{table*}
\centering
\caption{Summary of the imaging observations of 209P/LINEAR.}
\begin{tabular}{llccccccc}
\hline
Time (UT) & Facility & Res. & Exposure & Airmass & FWHM & $r_\mathrm{H}$ & $\varDelta$ & Plane Angle \\
 & & km/pix & min. & & arcsec & AU & AU & \\
\hline
2014 Apr 9.25 & GMOS-N & 23 & 0.2 & 1.7 & 0.4 & 1.043 & 0.441 & $-34.7^\circ$ \\
2014 May 18.75 & XM 0.35-m & 77 & 84 & 1.3 & 4.4 & 0.986 & 0.117 & $-16.9^\circ$ \\
2014 May 25.94 & F-2 & 8 & 10 & 1.7 & 0.8 & 1.009 & 0.064 & $+9.3^\circ$ \\
\hline
\end{tabular}
\label{tbl:obs}
\end{table*}

\clearpage

\begin{table*}
\centering
\caption{Input parameters for the Monte Carlo dust model. The orbital elements are extracted from the JPL elements 130, epoch 2011 Jun 8.0 UT.}
\begin{tabular}{ll}
\hline
Parameter & Value \\
\hline
Semimajor axis $a$ & 2.93102~AU \\
Eccentricity $e$ & 0.69237 \\
Inclination $i$ & $19.44783^\circ$ \\
Longitude of the ascending node $\Omega$ & $65.46431^\circ$ \\
Argument of perihelion $\omega$ & $150.46931^\circ$ \\
Epoch of perihelion passage $t_\mathrm{p}$ & 2009 Apr 17.43973 UT \\
Nucleus radius $R_\mathrm{N}$ & 1.35~km \\
Nucleus bulk density $\rho_\mathrm{N}$ & $500~\mathrm{kg \cdot m^{-3}}$ \\
Dust bulk density $\rho_\mathrm{d}$ & $1000~\mathrm{kg \cdot m^{-3}}$ \\
Dust albedo $A_\mathrm{d}$ & 0.05 \\
\hline
\end{tabular}
\label{tbl:mdl-orb}
\end{table*}

\clearpage

\begin{table*}
\centering
\caption{Dust model parameters derived from observations of Xingming 0.35-m (XM) and Gemini F-2 (F-2).}
\begin{tabular}{lll}
\hline
Parameter & Tested Values & Best-fit Values \\
\hline
Dust size lower limit, $\beta_\mathrm{rp,max}$ & 0.001--0.1 in steps of & XM: 0.005 \\
 & 1/40 of full range in log space & F-2: 0.004 \\
Mean speed of $\beta_\mathrm{rp}=1$ dust at 1~AU, $V_0$ & 10--400 in steps of 10 & XM \& F-2: $40\pm10~\mathrm{m \cdot s^{-1}}$ \\
Lagging parameter, $\sigma_\nu$ & 0.0--0.5 in steps of 0.1 & XM \& F-2: $0.3\pm0.1$ \\
Size index, $q$ & 2.6 to 4.4 with steps of 0.1 & XM \& F-2: $3.8\pm0.4$ \\
\hline
\end{tabular}
\label{tbl:mdl}
\end{table*}

\clearpage

\begin{table*}
\centering
\caption{Summary of the CMOR datasets used for analyzing the 2014 Camelopardalid outburst.}
\begin{tabular}{lll}
\hline
Label & $N$ & Description \\
\hline
\textit{Initial} & 99 & Extracted from processed daily data, \\
& & includes 85 underdense meteors and 14 overdense meteors. \\
\textit{Underdense} & 85 & Subset of \textit{initial} dataset, contains only underdense meteors. \\
\textit{Overdense} & 63 & Manually extracted from raw data, \\
& & includes 14 meteors from \textit{initial} dataset. \\
\hline
\end{tabular}
\label{tbl:met-dataset-summary}
\end{table*}

\end{CJK*}
\end{document}